\documentclass[twocolumn,prl,superscriptaddress,showpacs,preprintnumbers,amsmath,amssymb]{revtex4-1}
\usepackage{graphicx}
\usepackage{color}
\usepackage{ulem}
\usepackage{amssymb}
\usepackage{wasysym}

\begin{document}

\title{Specific heat in two-dimensional melting}

\author{Sven Deutschl\"ander}
\affiliation{Physics Department, University of Konstanz, 78464 Konstanz, Germany}
\author{Antonio M. Puertas}
\affiliation{Department of Applied Physics, University of Almeria, 04120 Almeria, Spain}
\author{Georg Maret}
\affiliation{Physics Department, University of Konstanz, 78464 Konstanz, Germany}
\author{Peter Keim}
\affiliation{Physics Department, University of Konstanz, 78464 Konstanz, Germany}
\date{\today}

\begin{abstract}
We report the specific heat $c_N$ around the melting transition(s) of micrometer-sized superparamagnetic particles confined in two dimensions, calculated from fluctuations of positions and internal energy, and corresponding Monte Carlo simulations. Since colloidal systems provide single particle resolution, they offer the unique possibility to compare the experimental temperatures of peak position of $c_N(T)$ and symmetry breaking, respectively. While order parameter correlation functions confirm the Kosterlitz-Thouless-Halperin-Nelson-Young melting scenario where translational and orientational order symmetries are broken at different temperatures with an intermediate so called hexatic phase, we observe a single peak of the specific heat within the hexatic phase, with excellent agreement between experiment and simulation. Thus, the peak is not associated with broken symmetries but can be explained with the total defect density, which correlates with the maximum increase of isolated dislocations. The absence of a latent heat strongly supports the continuous character of both transitions.
\end{abstract}

\maketitle

KTHNY theory, a microscopic melting scenario for two-dimensional solids developed by Kosterlitz, Thouless, Halperin, Nelson, and Young \cite{Kosterlitz1972Kosterlitz1973,Halperin1978Nelson1979,Young1979}, motivated extended analytic theories \cite{Fisher1979,Ramakrishnan1982,Chui1982,Kleinert1983,Kleinert19881989}, numerous experimental studies \cite{Widom1979,Brinkmann1982,Rosenbaum1983,Kusner1994,Zahn1999,Zahn2000,Gruenberg2004,Angelescu2005,Keim2007,Petukhov2005,Han2008,Brodin2010,Peng2010,Deutschlander2013,Horn2013,Schockmel2013} and simulations \cite{Frenkel1979,Strandburg1983,Udink1987,Chen1995,Lin2006,Shiba2009,Gribova2011,Prestipino2011,Wierschem2011,Toxvaerd1980,Abraham1980,Saito1982,Tobochnik1982,Strandburg1984,Weber1995,Marcus1996,Bernard2011,Prestipino2012,Dudalov2014,Loewen1996} to clarify the detailed melting mechanism and the order of phase transitions in 2D. The KTHNY melting is mediated by the dissociation of two kinds of topological defects, dislocations and disclinations. This scenario predicts two continuous phase transitions, where translational and orientational order is broken at different temperatures by the unbinding of pairs of dislocations and disclinations. In a triangular lattice, a disclination is a five- or sevenfold oriented site and a dislocation consists of oppositely charged disclinations, namely a pair of bound five- and sevenfold sites. Both types of topological defects can be treated as a Coulomb gas obeying a logarithmic interaction potential in two dimensions \cite{Minnhagen1987}. For the dislocation unbinding, a vector charge description becomes necessary due to the directional character of the defects, alongside a renormalization analysis accounting for their self-screening whereas a scalar charged gas of defects describes the disclination unbinding. Several experiments \cite{Kusner1994,Zahn1999,Zahn2000,Gruenberg2004,Keim2007,Han2008,Brodin2010,Peng2010,Deutschlander2013} and simulations \cite{Frenkel1979,Strandburg1983,Udink1987,Chen1995,Lin2006,Shiba2009,Gribova2011,Prestipino2011,Wierschem2011} clearly show the existence of the hexatic phase, but some studies additionally report first order characteristics \cite{Toxvaerd1980,Abraham1980,Tobochnik1982,Strandburg1984,Weber1995,Marcus1996,Angelescu2005} versus continuous order \cite{Keim2007,Han2008,Deutschlander2013}. Continuous and first-order characteristics have been observed within the same model, either differently for both KTHNY transitions \cite{Bernard2011} or preempted by a single first-order transition when the pair potential contains two length scales \cite{Prestipino2012,Dudalov2014}. It is suggested that the nature and number of transitions in 2D either depends on the dislocation core energy \cite{Chui1982,Saito1982} (which might implicitly depend on the particle pair interaction being short- or long-range) or the angular stiffness of the crystal being lower than a critical value \cite{Kleinert19881989}.

While first-order phase transitions are known to show a discontinuity in the free energy and a $\delta$-like divergence in the specific heat at the transition temperature, the defect free energy and specific heat of the two-dimensional Coulomb gas have only  discontinuities and no divergence for both transitions \cite{Kosterlitz1974,Halperin1978Nelson1979,Strandburg1988}. Thus, this feature can be used to identify the order of the transition. On the experimental side, there have been only calorimetric measurements so far, e.g. on atomic monolayers on graphite which show different results concerning the number of peaks in the specific heat, their position and magnitude \cite{Chung1979,Migone1984,Kim1986,Marx1988,Warken2000}. These experiments lack a precise determination of symmetry switching points, leaving the correlation to occurring phase transitions still elusive. Simulations of interacting dislocations show that for small dislocation core energies, the specific heat has a large discontinuity consistent with a first-order transition while for large core energies, a single moderate peak was observed, pointing to a continuous character \cite{Saito1982}. Laplacian roughening models \cite{Strandburg1983} which are dual to 2D melting and Lennard-Jones systems \cite{Wierschem2011} display one non-divergent peak along the two-step KTHNY scenario whereas a non-ideal Yukawa system shows two singularities associated with two transitions \cite{Vaulina2010}. However, in contrast to the atomic or molecular systems mentioned above, in colloidal systems, microscopy of individual particles allows a direct comparison between specific heat and symmetry switching points.

In this work, we present a melting study of superparamagnetic colloidal spheres confined in two dimensions and corresponding Monte Carlo simulations. The precise knowledge of the particle pair potential together with high precision single particle resolution and long term stability of the sample allows us to measure an anomaly in the specific heat in a colloidal system and compare it with simulations. Using order parameter correlation functions, we confirm in the experiments and simulations the two step KTHNY melting scenario from a solid phase through a hexatic fluid to an isotropic fluid, yet we find a single peak in the specific heat. Remarkably, this peak does not coincide with either transition temperature but lies within the hexatic phase. We show that it is connected to a sharp increase of the number of topological defects associated with a progressive unbinding of dislocation pairs on heating above the solid-hexatic transition temperature. Further, we do not find an additional peak correlated to the disclination unbinding which might not be resolvable due to the very small concentration of single disclinations $<5\permil$ in the background of a large overall defect density at the hexatic-isotropic transition.

The experimental system consists of an ensemble of spherical superparamagnetic polystyrene beads, with diameter $d=$4.5 $\mu$m and mass density 1.7 kg/dm$^3$, dissolved and sterically stabilized with sodium dodecyl sulfate in water. The colloidal suspension is sealed in a millimeter sized glass cell where sedimentation leads to the formation of a monolayer ($>10^5$ particles) on the bottom glass plate. The whole sample is under steady control and stable for more than 20 months which allows sufficient equilibration times and provides ideal sample conditions, e.g. vanishing density gradients or drifts. The ensemble is kept at room temperature and a highly homogeneous, finely tunable external magnetic field $H$ perpendicular to the colloidal layer induces a repulsive dipole-dipole interaction between the particles. This is quantified by the inverse system temperature that is defined as the ratio of the mean magnetic energy between two neighboring particles $E_\textrm{mag}$ and the thermal energy,
\begin{equation}\label{eq1}
 \Gamma=\frac{E_\textrm{mag}}{k_BT}=\frac{\mu_0\left(\pi n\right)^{3/2}\left(\chi H\right)^2}{4\pi k_BT}
\end{equation}
where $n=1/a_0^2$ is the 2D particle density with the mean particle distance $a_0$, and $\chi$ the magnetic susceptibility of the beads. We assume an error of $\Gamma\pm0.5$ due to density and room temperature fluctuations during the measurements. After changing the interaction strength, the system is equilibrated for at least 24 hours before $\approx$ 3000 particles are monitored and tracked by video microscopy. Previous studies of this system have shown excellent agreement with the KTHNY melting scenario \cite{Zahn1999,Zahn2000,Gruenberg2004,Keim2007,Deutschlander2013,Horn2013}.
\begin{figure}[ht]
\centering
\begin{tabular}{c}
\includegraphics[width=0.95\linewidth]{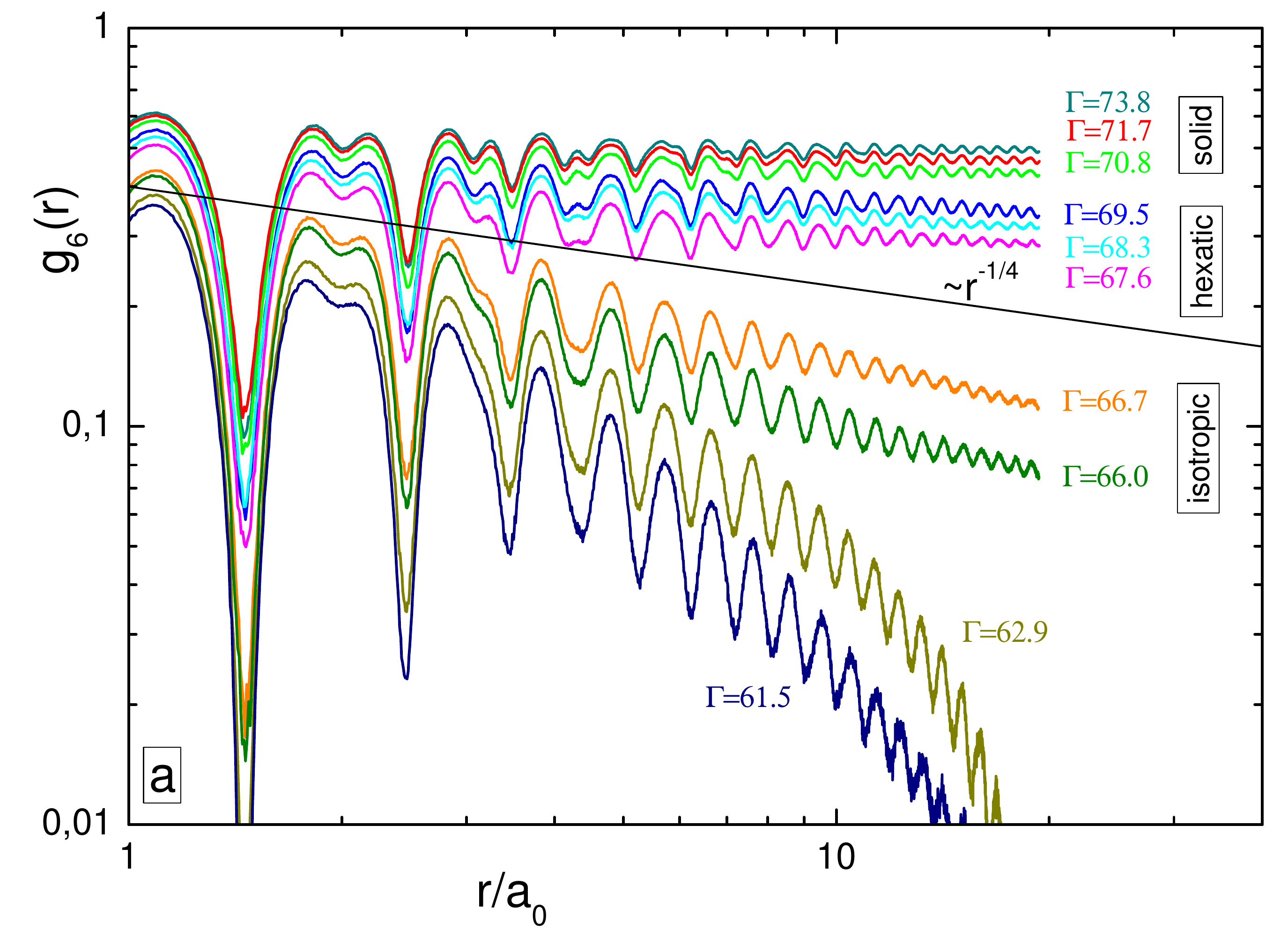} \\
\includegraphics[width=0.95\linewidth]{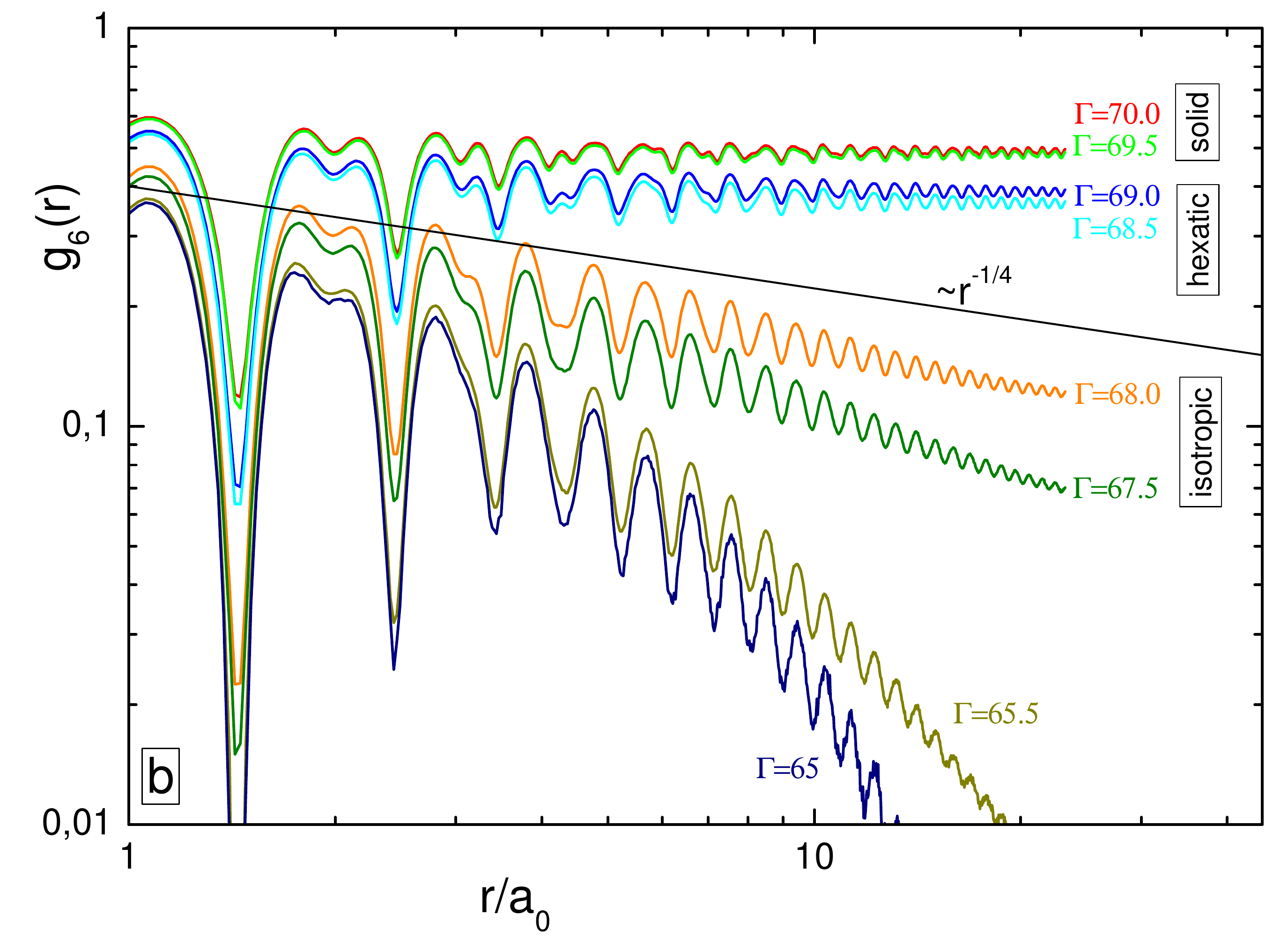}
\end{tabular}
\caption{(Color online) Spatial orientational correlation $g_6(r)=\left\langle \psi_6^*(r)\psi_6(0)\right\rangle$ at different system temperatures $\Gamma$ for experiment (a) and simulation (b). The data are plotted on a log-log scale in reduced coordinates, where $a_0=(n)^{-1/2}$ is the mean particle distance in the respective system . The decay behavior of $g_6(r)$ has distinct characteristics in the solid (constant), the hexatic liquid (algebraic decay) and in the isotropic liquid (exponential decay). An algebraic exponent of $-1/4$ marks the hexatic-isotropic transition \cite{Halperin1978Nelson1979}.}
\label{fig1}
\end{figure}

In addition, standard Monte Carlo simulations are run in the NVT ensemble, with $N=2500$ particles interacting with a dipolar potential: $\beta V(r)=\Gamma / r^3$, with distances measured in units of $(\pi n)^{-1/2}$. The interactions are cut off at $R_{cut}=9a_0$, which is large enough to avoid effects from the truncation \cite{Lin2006}. The system is simulated in a rectangular box with a size ratio $2:\sqrt{3}$ and a (hard disc) 2D area fraction $\phi=0.07$ to mimic the experimental conditions. Cycles of increasing $\Gamma$ from the fluid to the crystal, and subsequent decrease to the fluid again were used to confirm that there is no hysteresis within our $\Gamma$-resolution.

To determine the respective symmetry breaking temperatures, we analyze the spatial correlation function $g_6(r)=\left\langle \psi_6^*(r)\psi_6(0)\right\rangle$ of the orientational order parameter $\psi_{6}=\frac{1}{n_j}\sum_{k}\exp\left(i6\theta_{jk}\right)$, where the sum runs over all $n_j$ nearest neighbors of particle $j$, and $\theta_{jk}$ is the angle of the $k$-th bond with respect to a certain reference axis. According to the KTHNY theory, $g_6(r)$ approaches a constant value in the solid phase (long-range order), decays algebraically $\sim r^{-\eta_6}$ in the hexatic fluid (quasi-long-range) and exponentially $\sim e^{-r/\xi_6}$ in the isotropic fluid (short-range), with an orientational correlation length $\xi_6$. The orientational exponent $\eta_6$ is inversely proportional to the orientational stiffness of the system: Infinite in the solid, zero in the isotropic fluid and finite in the hexatic phase, approaching a value of $\eta_6\sim 1/4$ at the hexatic-isotropic transition \cite{Halperin1978Nelson1979,Keim2007}. This behavior of $g_6(r)$ has successfully been probed and verified in experiments \cite{Zahn1999,Keim2007,Han2008,Brodin2010,Peng2010} and simulations \cite{Udink1987,Lin2006,Shiba2009,Gribova2011,Prestipino2011}. The results for our system are shown in Fig.~\ref{fig1}: for both experiment and simulation, we clearly observe the characteristic behavior of $g_6(r)$ for the different phases alongside equidistant $\Gamma$-steps, verifying the stability of the orientational quasi-long-range ordered hexatic fluid. For the solid-hexatic transition, we find the (inverse) transition temperatures  $\Gamma_m^{exp}\approx70.3$ and $\Gamma_m^{sim}=69.25$, for the hexatic-isotropic transition we find $\Gamma_i^{exp}\approx67.3$ and $\Gamma_i^{sim}=68.25$ \cite{Transitions}. These values are extracted from Fig.~\ref{fig1}. Since the solid-hexatic transition is more difficult to locate by $g_6(r)$, we present a finite-size analysis of the translational order parameter in the supplemental material confirming these values. It is well known that the width of the hexatic phase is affected by the system size which might explain the different melting temperatures in experiment and simulation.
\begin{figure}[ht!]
\centering
\includegraphics[width=0.98\linewidth]{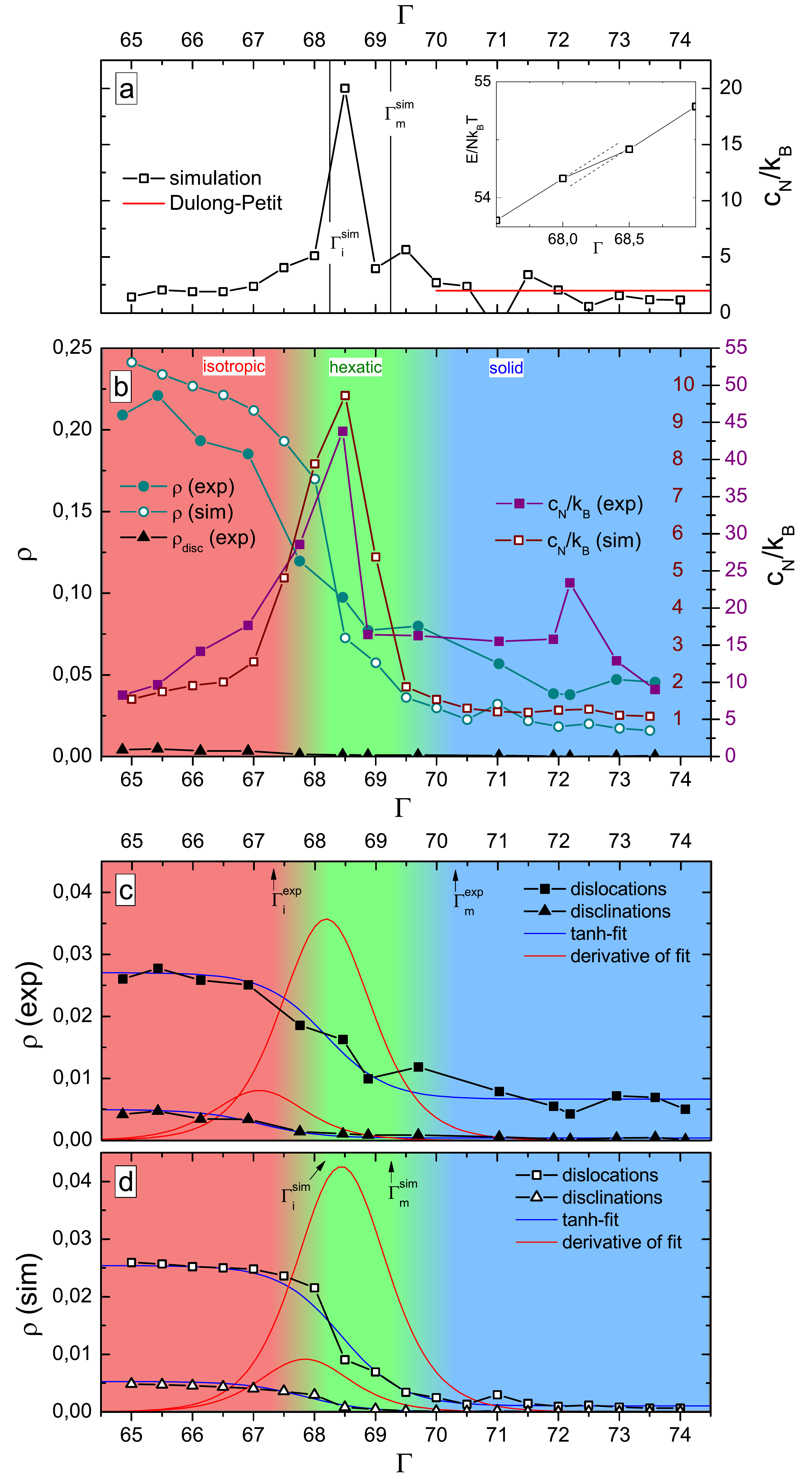}
\caption{(Color online) Filled symbols correspond to experiment, open symbols to simulation. (a): Specific heat $c_N/k_B$ as a function of $\Gamma$ calculated via the derivative approach. The inset shows the energy per particle. (b): $c_N/k_B$  for experiment (right side of the right scale) and simulation (left side of the right scale) from energy fluctuations and the total defect number density (left scale), counting all defects ($\rho$) and disclinations ($\rho_{\textrm{disc}}$). (c and d): The behavior of isolated dislocations and disclinations analyzed with tanh-fit to show the steepest increase. The background color of (b,c,d) is organized as follows: The solid and isotropic fluid found by the experiment are colored blue and red (dark grey), respectively. The hexatic phase is colored green (light grey). The color gradient shows the estimated error in determining the transition temperatures.}
\label{fig2}
\end{figure}

According to the KTHNY theory, the specific heat at constant pressure $C_p$ behaves as $\xi^{-2}$ for both transitions, with the orientational correlation length $\xi_6$ from the isotropic fluid to the hexatic phase at $T_i$, and the translational correlation length $\xi_T$ at the transition from the hexatic to the solid phase at $T_m$, respectively \cite{Kosterlitz1974,Halperin1978Nelson1979}. For the unbinding of dislocation pairs e.g. it reads $C_p\sim\exp\left(-b/|T-T_m|^{\bar{\nu}}\right)$, where $b$ is a constant and $\bar{\nu} = 0.36963$ is the critical exponent from renormalization group theory \cite{Halperin1978Nelson1979,Young1979}. At the transition points, the specific heat undergoes only an essential singularity and no divergence. However, a system might show a weak peak above $T_m$ (below $\Gamma_M$) caused by a successive unbinding of dislocation pairs while position, width and height of the peak strongly depend on the model \cite{Strandburg1988}. The specific heat $c_N$ per particle and at constant volume can be calculated via the derivative of the internal energy with respect to temperature (inverse $\Gamma$) or from energy fluctuations (see supplemental material),
\begin{equation}
\label{eq6}
c_N=\frac{1}{N}\frac{\partial\left\langle E\right\rangle}{\partial T}=-\frac{k_B\Gamma^2}{N} \frac{d (\langle E\rangle/\Gamma)}{d\Gamma}=
\frac{\left\langle E^2\right\rangle-\left\langle E\right\rangle^2}{Nk_BT^2}
\end{equation}
where $E$ is the total internal energy of the $N$-particle system and the brackets denote a time average \cite{Cn}. The results are shown in Fig.~\ref{fig2}: for the energy summation, the cutoff is set to $15a_0$ for the experiment and $9a_0$ for the simulation (further discussion of the cutoff dependency in the supplemental material). Within the simulations, the calculation of $c_N$ from the derivative of the energy (Fig.~\ref{fig2}a) and its fluctuations (Fig.~\ref{fig2}b) agree almost quantitatively and show a single peak at $\Gamma^{sim}_{c_N}=68.5$ due to a single change of slope in the energy (inset). In the solid phase we observe a value for $c_N$ in agreement with the Dulong-Petit law that predicts the heat capacity of a two-dimensional monatomic crystal in the harmonic approximation, $C_V=2Nk_B$ (horizontal line in Fig.~\ref{fig2}a). For the experiments, the calculation from the derivative of the energy is too noisy (see supplemental material), and a reliable value can be obtained only from the energy fluctuations (Fig.~\ref{fig2}b). We find again a single marginal peak at $\Gamma^{exp}_{c_N}\approx68.4$, very close to the value of the simulations.
(Note, however, the different scale for experiment and simulation of $c_N$. As discussed in the supplemental material, increased peak height and baseline can be attributed to additional density fluctuations picked up in the experiment which, nevertheless, do not affect the peak position.) The (inverse) temperature $\Gamma_{c_N}$ of the specific heat peak lies within the hexatic phase below the melting temperature $\Gamma_m$ from solid to hexatic ($\Gamma_{c_N}<\Gamma_m$ or $T_{c_N}>T_m$), both in the experiments and simulations. A second peak is not detectable unexpectedly, given the picture of the KTHNY theory which predicts two specific heat discontinuities (eventually marginal) but located at the transition temperatures \cite{Halperin1978Nelson1979}. Already in the 1950's, a shift of the specific heat peak in 2D systems has been reported for quantum fluids like $^4$He films, whose position is found at higher temperatures with respect to the onset of superfluidity \cite{Brewer1965,Symonds1966}. The authors put this on the increasing importance of surface excitations with reduced film thickness. De Gennes (comment in \cite{Symonds1966}) pointed out that the temperature onset of superfluidity might be caused by short-range-order effects which become important in one- and two-dimensional salts \cite{Marel1955,Gijsman1959}. Later, Kosterlitz and Thouless considered this effect for the neutral superfluid in 2D \cite{Kosterlitz1972Kosterlitz1973}. Berker and Nelson gave analytic evidence for a specific heat shift for superfluid films of $^3$He-$^4$He mixtures, and explained this with the gradual unbinding of vortex pairs with increasing temperature while the maximum in the specific heat occurs when the mean separation of vortex pairs is comparable with the vortex core size \cite{Berker1979}.  A shift of the specific heat peak to higher temperatures has also been reported in simulations of planar models \cite{Tobochnik1979,Himbergen1981,Solla1981} and 2D solids \cite{Saito1982,Strandburg1983,Wierschem2011}.

To explain the shift in the specific heat singularity, we investigate local quantities, in particular, the defect distributions. We analyzed the total number density $\rho$ for all defects (not sixfold coordinated sites, Fig.~\ref{fig2}b) as well as the density of isolated dislocations (one dislocation is counted twice, containing two defects) and disclinations, for experiment (Fig.~\ref{fig2}c) and simulation (Fig.~\ref{fig2}d). We find that in the region of the specific heat peak the overall defect density $\rho$ undergoes a significant increase from $\approx5\%$ to $\approx20\%$. Energy costs which become apparent in the specific heat should directly be connected to the creation and dissociation of defects, and should peak where the increase of defects is large which is not necessarily at $\Gamma_m$. This can clearly be seen from the simulations where the sharpest increase of $\rho$ is exactly at the peak position $\Gamma^{sim}_{c_N}=68.5$ $(\Delta\rho\approx0.1)$. With the total defect increase $\Delta N_{def}\approx220$ at this interaction strength, we can make a rough estimate for the simulation peak height via the the dislocation core energy in the hexatic phase $E_c\approx5.5k_BT$~\cite{Eisenmann2005} observing $c_N\approx30k_B$ (more detailed in the supplemental material). The defect density in the experiment on the other hand, shows a rather broad increase. It must be noted that this includes all kinds of defects, including cluster conformations which earliest occur in the hexatic phase. Such clustering is beyond KTHNY theory which assumes a dilute gas of defects but is quite natural due to the attractive interaction of the defects. Implicitly, we can extract from the larger specific heat peak height estimated by the total defect increase, compared to the derivative or the fluctuation of the internal energy, that such clustered dislocations have a significantly reduced core energy $E_c<5.5k_BT$. We checked that such cluster consists only of an equal amount of five- and seven-folded particles in the hexatic phase: clusters are dislocation-cluster (with small core energy) but not dislocation-disclination cluster. The latter, with unequal number of five- and seven-folded particles are only observable quite deep in the isotropic phase (see supplemental material). Thus, we focus on the isolated topological defects that drive the transitions within KTHNY theory: isolated dislocations show their sharpest increase in density below $\Gamma_m$ but very close to the peak position ($\Gamma_{disl}^{exp}\approx68.2$ for the experiment and $\Gamma_{disl}^{sim}\approx68.4$ for the simulation). On the other hand, the increase of isolated disclinations is as well shifted in respect to the hexatic-isotropic transition, but only marginally. We do not observe an indication of a second specific heat peak corresponding to this unbinding because this implies an increase of less than $5\permil$ in the background of a large overall defect density (see Fig.~\ref{fig2}b) and a negligible increase of the energy: a rough estimate of the peak height as above but due to disclination unbinding, disclination core energy ($\approx5k_BT$) times number of unbinding disclinations, gives $\approx1 k_B$.

Using a colloidal model system and Monte Carlo simulations, we measure the specific heat via fluctuations of the internal energy. We observe a single peak in the specific heat above the solid-hexatic transition ($T_{c_N}>T_m$), although melting in 2D shows two phase transitions at distinct temperatures. The peak in $c_N$ arises when the change in the defect density is largest, what appears within the hexatic phase and not directly at $T_m$. Whereas only a few defects are needed to destroy the given order, their cost in energy is small at $T_m$ (and $T_i$). A second peak in $c_N$ associated to disclination unbinding from dislocations is not detectable since their number density stays small compared to the overall defect density even deep in the isotropic fluid phase. We can further conclude that the absence of a latent heat strongly supports the continuous character of both transitions.

\begin{acknowledgments}
A.M.P. acknowledges financial support from the Consejera de Econom\'{\i}a, Innovaci\'on y Ciencia (Junta de Andaluc\'{\i}a) and from the Ministerio de Ciencia e Innovaci\'on, and FEDER funds, under projects P09-FQM-4938 and MAT2011-28385, respectively. P.K. gratefully acknowledges financial support from the German Research Foundation (DFG), project KE 1168/8-1.
\end{acknowledgments}

\vspace{-0.5cm}
\bibliographystyle{prsty}

\begin{thebibliography}{10}
\bibitem{Cn} Since we have a purely repulsive system with algebraic distance dependence, we have only a single control parameter which is $\Gamma$. For a given particle number, pressure ($\sim\chi^2H^2 $) and volume cannot be changed independently. Therefore we denote the specific heat of this particular system with the subscript $'N'$.
\end{thebibliography}

\newpage
\begin{figure*}
\centering
\includegraphics[width=\linewidth]{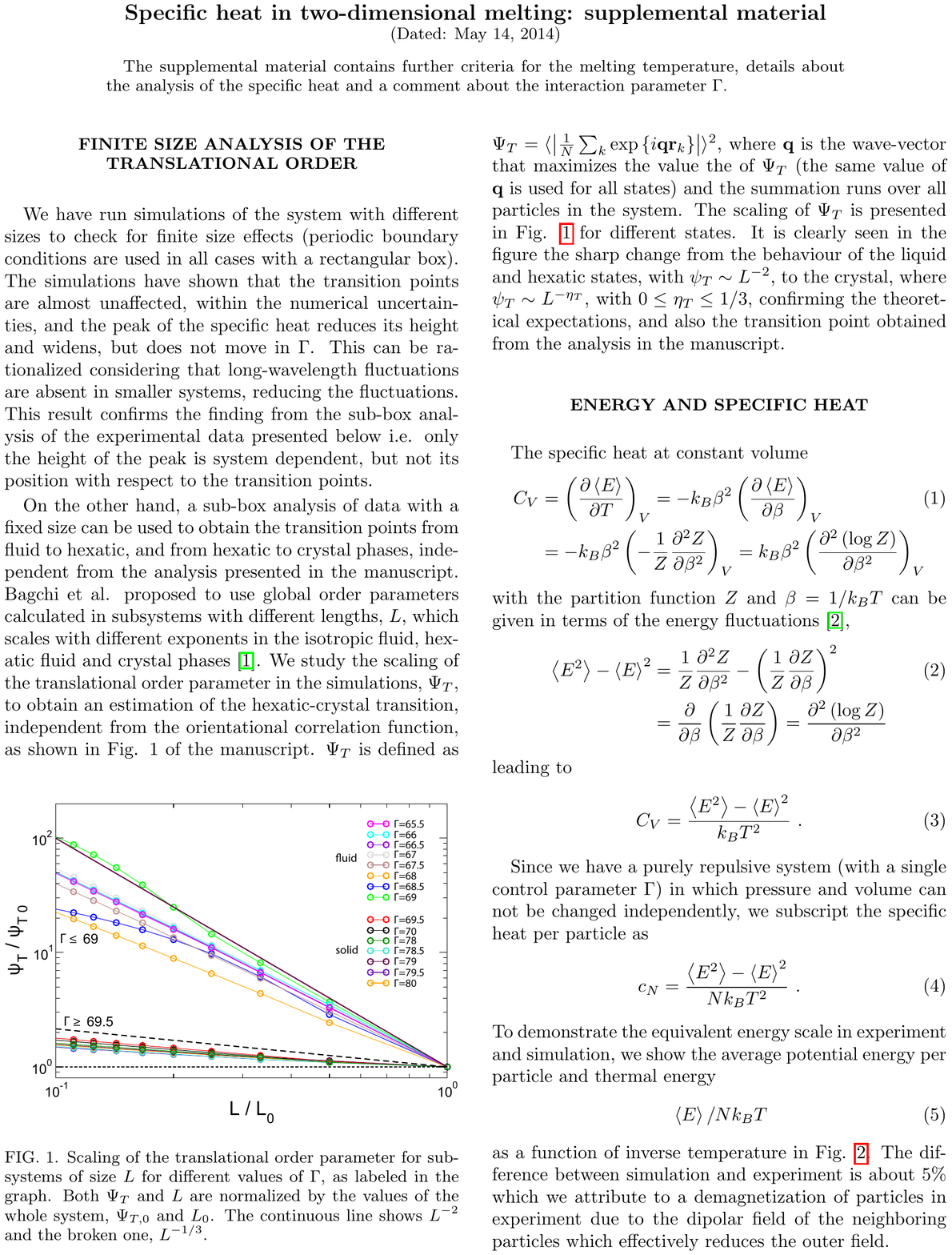}
\end{figure*}
\newpage
\begin{figure*}
\centering
\includegraphics[width=\linewidth]{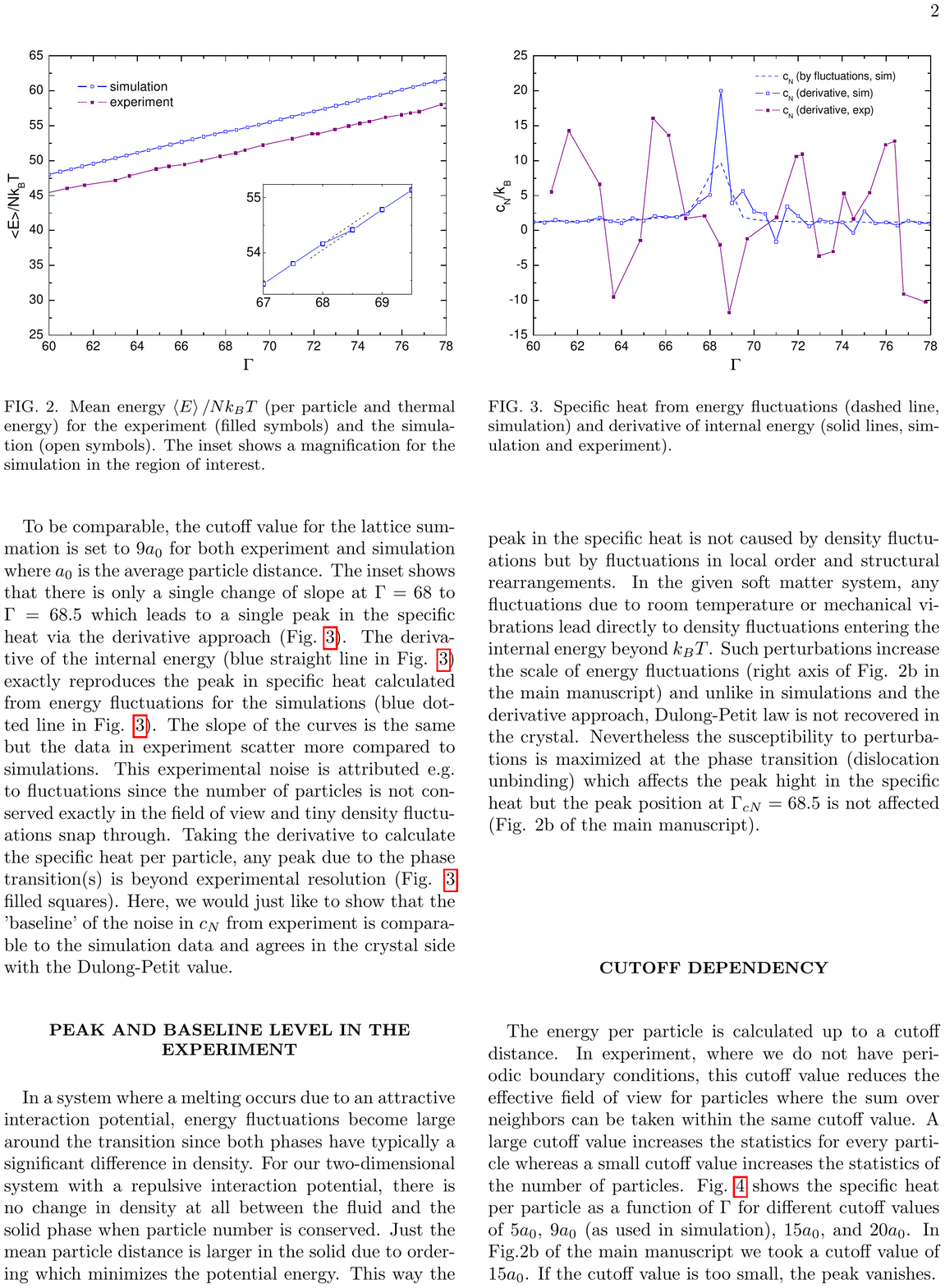}
\end{figure*}
\newpage
\begin{figure*}
\centering
\includegraphics[width=\linewidth]{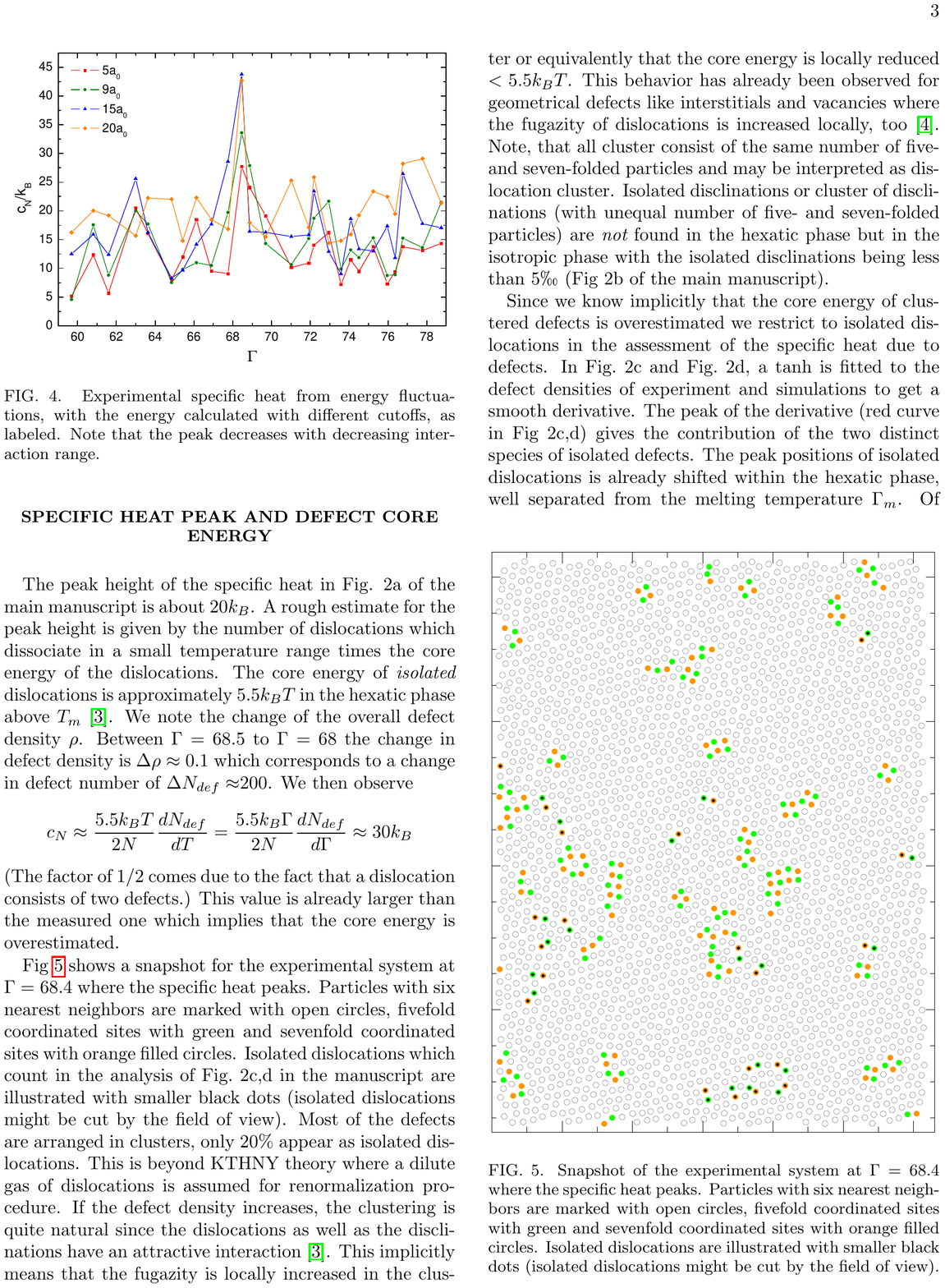}
\end{figure*}
\newpage
\begin{figure*}
\centering
\includegraphics[width=\linewidth]{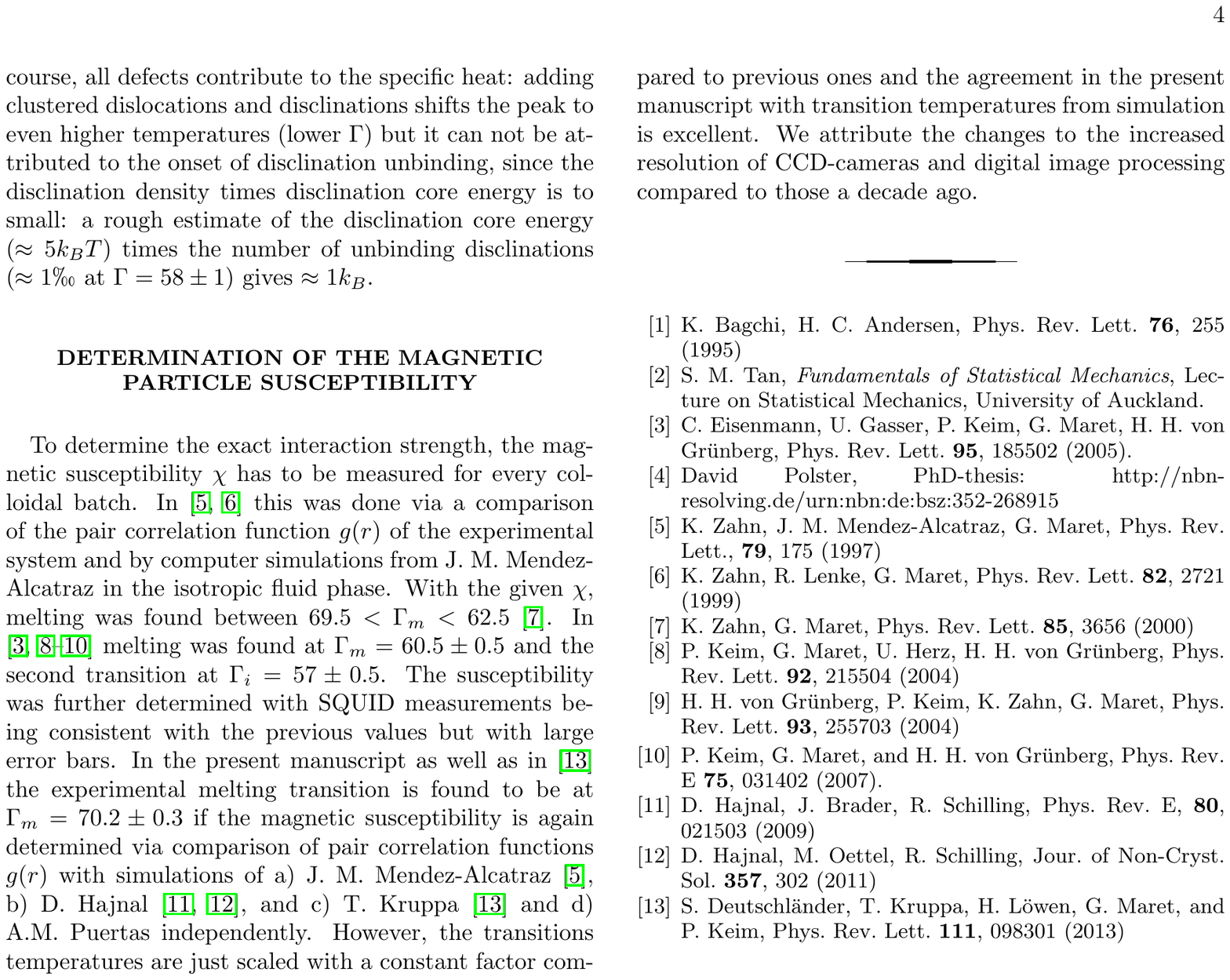}
\end{figure*}

\end{document}